\documentclass[smallextended]{svjour3m}

\usepackage{graphicx}

\usepackage{hyperref}

\journalname{Gen Relativ Gravit}

\begin{document}

\title{On the possibility of observation of the future for movement
in the field of black holes of different types}

\titlerunning{On the possibility of observation}

\author{Yuri V. Pavlov}

\institute{Yu. V. Pavlov \at
              Institute of Problems in Mechanical Engineering,
Russian Academy of Sciences,\\
Bol'shoy pr. 61, V.O., Saint Petersburg 199178, Russia\\
              \email{yuri.pavlov@mail.ru}
}

\maketitle

\begin{abstract}
    It is shown for a spherically symmetric black hole of general type that
it is impossible to observe the infinite future of the Universe
external to the hole during the finite proper time interval of the free fall.
    Quantitative evaluations of the effect of time dilatation for circular
orbits around the Kerr black hole are obtained and it is shown that the effect
is essential for ultrarelativistic energies of the rotating particle.

\end{abstract}

\section{Introduction}
\label{intro}
    It is well known an observer falling radially into a black hole will
reach the horizon in finite proper time, but the coordinate time in
the Schwarzschild coordinate system is infinite~\cite{LL_II,MTW}.
    This leads to an illusion of the possibility for a falling in the black
hole by a cosmonaut so as to observe the infinite future of
the Universe external to the black hole~(see,
for example,~\cite{Redze,Cherepashchuk}).
    The impossibility of such an observation is shown
in~\cite{GribPavlov2008UFN,Krasnikov08}.
    Note that the impossibility of the observation of the infinite future
for the radial falling to the Schwarzschild black hole in the
four-dimensional space-time is evident from the properties of the
Kruskal--Szekeres coordinate system~\cite{Kruskal60,Szekeres60}.

    The possibility of observing the infinite future when falling on the black
hole is analyzed in Sect.\,\ref{sec:1}
for spherically symmetrical black holes of general type:
black holes with electrical charge,
with nonzero cosmological constant, dirty black holes
(those with nonzero stress energy outside of static horizons),
and multidimensional black holes.
    In the general case, there are no explicit analytical expressions for
Kruskal--Szekeres coordinates and the analysis of space-time properties in
such cases is difficult even for the radial falling on a black hole.
    In this paper it is shown from the analysis of the null and time-like
geodesics for spherically symmetric black holes of the general type
that, if the proper time of the fall is finite, then the time interval of
observation of events in the point of the beginning of the fall is
also finite.
    Quantitative evaluations are given for the Schwarzschild black holes.

    Another the possibility to observe the far future of the external Universe
is also discussed in the literature, namely due to time dilatation near
a black hole.
    So in~\cite{Rees}, p.~92, it is stated
``However, a more prudent astronaut who managed to
get into the closest possible orbit around a rapidly spinning
hole without falling into it would also have interesting
experiences: space-time is so distorted there that his clock would
run arbitrary slow and he could, therefore, in subjectively
short period, view an immensely long future timespan in
the external universe''.
    In Sect.\,\ref{sec:2} of the paper quantitative evaluations for the
time dilatation on the circular orbits around the rotating black hole
are obtained and it is shown that the effect becomes essential for
ultrarelativistic energies of the rotating object.

\section{
Observation of the future when falling on the spherically symmetric black
holes}
\label{sec:1}

    Consider a spherically symmetric black hole with the metric
        \begin{equation} \label{NB11}
ds^2 = A(r) \, c^2 dt^2 - \frac{dr^2}{A(r)} - r^2 d \Omega_{N-2}^{\,2} \,,
\end{equation}
    where $c$ is the light velocity, $d \Omega_{N-2}$ --- the angle element
in space-time of the dimension $N \ge 4$,
$A(r)$ --- a certain function of the radial coordinate~$r$
which is zero on the event horizon~$r_H$ of the black hole: $A(r_H)=0$.
    For the Schwarzschild black hole it holds~\cite{Schwarzschild16}
    \begin{equation} \label{MBHo}
A(r) = 1 - \frac{r_g}{r} \,, \ \ \ \ r_g=\frac{2GM}{c^2} \,,
\end{equation}
    where $G$ is the gravitational constant,
$M$ --- the mass of the black hole.
    For an electrically charged nonrotating black hole in
vacuum we have~\cite{Reissner16,Nordstrom18}
    \begin{equation} \label{MBHo1}
A(r) = 1 - \frac{r_g}{r} + \frac{q^2}{r^2}\,,
\end{equation}
    where $q$ is the charge of the black hole.
    For nonrotating black holes with nonzero cosmological constant~$\Lambda$,
one has Kottler~\cite{Kottler18} solution
    \begin{equation} \label{MBHo2}
A(r) = 1 - \frac{r_g}{r} - \frac{\Lambda r^2 }{3} \,.
\end{equation}
    For multidimensional nonrotating charged black holes~\cite{Tangherlini63}
with a cosmological constant, one has
    \begin{equation} \label{MBHo3}
A(r) = 1 - \left( \frac{r_g}{r} \right)^{\!\! N\!-\!3} \!\!\!\!-
\frac{2 \Lambda r^2 }{(N \!-\! 1)(N\!-\!2)} +
\frac{q^2}{r^{2(N \!-\!3)}},
\end{equation}
    where
    \begin{equation} \label{MBH6}
r_g \stackrel{\rm def \mathstrut}{=}
\left[ \frac{ 2 G_N M}{(N-3) c^2} \right]^{1/(N-3)},
\end{equation}
    provided $G_N$ --- the $N$-dimensional gravitational constant is normalized
so that the $N$-dimensional Newton law in non-relativistic approximation
possesses the following form
    \begin{equation} \label{NNewton}
F = G_N \frac{mM}{r^{N-2}} \,.
\end{equation}

    Equations for geodesics in metric~(\ref{NB11}) can be written as
    \begin{equation} \label{NSF1}
A(r) \frac{dt}{d \tau} = \varepsilon , \ \ \ \
\frac{d \varphi}{ c d \tau} = \frac{L}{r^2} ,
\end{equation}
    \begin{equation} \label{NSF2}
\left( \frac{dr}{c d \tau} \right)^{\! 2} = \varepsilon^2 -
A(r) \left( \kappa + \frac{L^2}{r^2} \right),
\end{equation}
    where $\kappa = 1 $ is for timelike geodesics
and $\kappa = 0 $ is for the null geodesics.
    For a particle with the rest mass~$m$, the parameter $\tau$ is the proper
time, $\varepsilon m c^2 = {\rm const} $ is its energy in the gravitational
field~(\ref{NB11}); and $ L m c = {\rm const} $ is the projection of the
angular momentum on the axis orthogonal to the plane of movement
in the four-dimensional case.

    From Eqs.~(\ref{NSF1}), (\ref{NSF2}) for the intervals of the coordinate
time $ t_f - t_0 $ and proper $ \Delta \tau $ time of movement of the particle
from the point with the radial coordinate~$r_0$ to the point with
coordinate $r_f<r_0$, one has
    \begin{equation} \label{NSF3}
t_f - t_0 = \frac{1}{c} \int \limits_{r_f}^{r_0} \!
\frac{d r}{ \displaystyle A(r) \sqrt{ 1 - \frac{A(r)}{\varepsilon^2}
\left( \kappa + \frac{L^2}{r^2} \right)}}\,,
\end{equation}
    \begin{equation} \label{NSF4}
\Delta \tau = \frac{1}{c} \int \limits_{r_f}^{r_0} \!
\frac{d r}{ \displaystyle
\sqrt{ \varepsilon^2 - A(r) \left( 1 + \frac{L^2}{r^2} \right)}}\,.
\end{equation}
    As one can see from~(\ref{NSF3}), the smallest coordinate time of movement
is realized for photons with zero angular momentum.
    It is equal to
    \begin{equation} \label{NSF5}
t_f - t_s = \frac{1}{c} \int \limits_{r_f}^{r_0} \!
\frac{d r}{ A(r) }\,,
\end{equation}
    where $t_s$ is the starting time for radial movement of the photon
from the point~$r_0$.

    Subtracting~(\ref{NSF5}) from~(\ref{NSF3}) for $\kappa=1$, one finds
an answer to the question:
how much later are the events in points with the same value of the radial
coordinate as in the beginning of the fall which can be observed by the
observer falling up to the point~$r_f $?
    \begin{equation} \label{NB17dd}
t_s - t_0 = \frac{1}{c}
\int \limits_{r_f}^{r_0} \!\!\!
\frac{\frac{1}{\varepsilon^2} \left( 1 + \frac{L^2}{r^2}  \right) dr}
{\sqrt{ 1 \!-\! \frac{A(r)}{\varepsilon^2}
\Bigl( 1 \!+\! \frac{L^2}{r^2} \Bigl) }
\left[ 1 \!+\! \sqrt{ 1 \!-\! \frac{A(r)}{\varepsilon^2}
\Bigl( 1 \!+\! \frac{L^2}{r^2} \Bigr) } \right]}.
\end{equation}
    From~(\ref{NSF4}) and (\ref{NB17dd}), one arrives at the following
conclusion:
if the proper time interval is finite, then the time interval of observation
of the future events at the point of the beginning of the fall in the process
of falling is also finite.

    This conclusion is generalized for the case of movement of the charged
particle.
    For the metric~(\ref{MBHo3}) this can be obtained by the transformation
$ \varepsilon \to \varepsilon -(qQ/r^{N-3})$, where $Q$ is the specific
charge of the moving particle in
Eqs.~(\ref{NSF1})--(\ref{NSF4}), (\ref{NB17dd}).
    If the energy, angular momentum and the charge of the particle are such
that the proper time of the fall on the black hole is finite, then the
observation of the infinite future of the external Universe is impossible.

\begin{figure}[ht]
\centering
    \includegraphics[width=77mm]{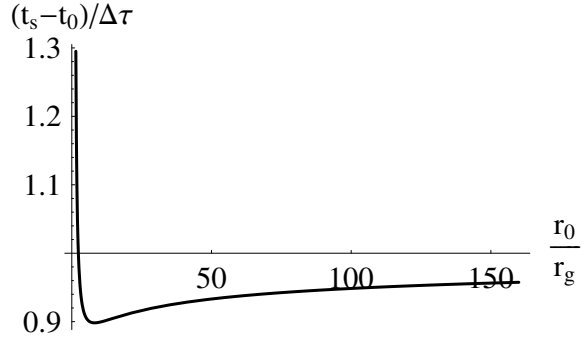}
\caption{The ratio $ (t_s - t_0) / \Delta \tau$ for the observer falling
from rest at the point $r_0$ to the horizon of the Schwarzschild black hole.
\label{DtDtau}}
\end{figure}

    In Fig.~\ref{DtDtau} the results of calculations of the ratio of
the possible time of observation of the future at the point of the fall
to the proper time interval for the observer radially falling in the
Schwarzschild black hole are given.
    As we can see from~(\ref{NSF4}), (\ref{NB17dd}) for the Schwarzschild
black hole (see explicit formulas (6), (9) from~\cite{GribPavlov2008UFN})
the following asymptotic behaviour holds
$ (t_s - t_0) / \Delta \tau \to 1$ for $r_0 \to \infty$.
    If the fall begins from rest at the point close to the event horizon,
then
    \begin{equation} \label{rtorg}
\frac{t_s - t_0}{\Delta \tau} \sim \sqrt{\frac{r_g}{r_0 - r_g}}\, \log 2
\to \infty, \ \ \ \ r_0 \to r_g.
\end{equation}
    But in this case $ t_s - t_0 \approx  (r_g/c)\, 2 \log 2$,
i.e., the possible time interval of the observation of the future is small.

    For any nonradial fall of the nonrelativistic particle on the Schwarzschild
black hole it is also impossible to have a large interval of the future time
which can be seen from Fig.~\ref{DDttL}:
    \begin{figure}[ht]
\centering
   \includegraphics[width=77mm]{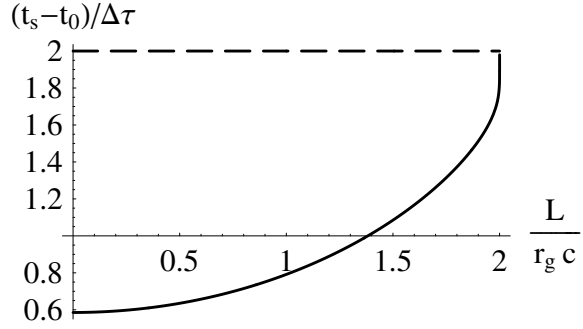}
\caption{The dependence $ (t_s - t_0) / \Delta \tau$ on the angular momentum of
the particle with $\varepsilon=1$, \, $r_0=3 r_g$ falling to the horizon of
the Schwarzschild black hole.
\label{DDttL}}
\end{figure}

\section{Time dilatation on circular orbits of the Kerr black hole}
\label{sec:2}

    Kerr's metric~\cite{Kerr63} of the rotating black hole in
Boyer-Lindquist~\cite{BoyerLindquist67} coordinates has the form
    \begin{eqnarray}
d s^2 = d t^2 \!-
\frac{2 M r \, ( d t \! - a \sin^2 \! \theta\, d \varphi )^2}{r^2 + a^2 \cos^2
\! \theta } - (a^2 \!\cos^2 \! \theta
\nonumber \\
+\, r^2 ) \Bigl( \frac{d r^2}{\Delta} + d \theta^2 \Bigr)
- (r^2 \!+ a^2) \sin^2 \! \theta\, d \varphi^2,
\label{Kerr}
\end{eqnarray}
    where
    \begin{equation} \label{Delta}
\Delta = r^2 - 2 M r + a^2,
\end{equation}
    $M$ is the mass of the black hole, $aM$ --- its angular momentum.
    Here we use the units: $c=G=1$.
    For $a=0$, the metric~(\ref{Kerr}) describes a nonrotating black hole in
Schwarzschild coordinates.
    The event horizon of the Kerr's black hole corresponds to the radial
coordinate
    \begin{equation}
r = r_H \equiv M + \sqrt{M^2 - a^2} \,.
\label{Hor}
\end{equation}

    Equatorial $(\theta=\pi/2)$ geodesics in Kerr's metric~(\ref{Kerr})
are defined by the equations (see~\cite{Chandrasekhar}, Sect.\,61):
    \begin{equation} \label{geodKerr1}
\frac{d t}{d \tau} = \frac{1}{\Delta} \left[ \left(
r^2 + a^2 + \frac{2 M a^2}{r} \right) \varepsilon - \frac{2 M a}{r} L \right],
\end{equation}
    \begin{equation}
\frac{d \varphi}{d \tau} = \frac{1}{\Delta} \left[ \frac{2 M a}{r}\,
\varepsilon + \left( 1 - \frac{2 M}{r} \right)\! L \right],
\label{geodKerr2}
\end{equation}
    \begin{equation} \label{geodKerr3}
\left( \frac{d r}{d \tau} \right)^2 = \varepsilon^2 +
\frac{2 M}{r^3} \, (a \varepsilon - L)^2 +
\frac{a^2 \varepsilon^2 - L^2}{r^2} - \frac{\Delta}{r^2} \kappa ,
\end{equation}
    where $\varepsilon m = {\rm const} $ is the energy of the particle
with the rest mass~$m$ in the gravitational field~(\ref{Kerr});
     $ L m = {\rm const} $ is the projection of the angular momentum of
the particle on the rotation axis of the black hole.

    Let us define the effective potential of the particle in the field of
the black hole by
    \begin{equation} \label{Leffdef}
V_{\rm eff} = -  \frac{1}{2} \left( \frac{d r}{d \tau} \right)^{\!2}.
\end{equation}
    Then $ d^2 r / d \tau^2 = - d V_{\rm eff} / d r $
and the necessary conditions for the existence of circular orbits in
equatorial plane are
    \begin{equation} \label{LeffCucl}
V_{\rm eff}=0, \ \ \ \ \ \ \ \frac{d V_{\rm eff}}{d r} =0\,.
\end{equation}
    It is sufficient for the existence of stable circular orbits that
    \begin{equation} \label{LeffCuclUst}
V_{\rm eff}=0, \ \ \ \ \frac{d V_{\rm eff}}{d r} =0, \ \ \ \
\frac{d^2 V_{\rm eff}}{d r^2} >0 \,.
\end{equation}

    The solutions of Eqs.~(\ref{LeffCucl}) can be written in
the form~\cite{BardeenPressTeukolsky72}
    \begin{equation} \label{ShapTyu1a}
\varepsilon = \frac{x^{3/2} - 2 \sqrt{x} \pm A }
{\sqrt{x \left( x^2- 3 x \pm 2 A \sqrt{x}\, \right) }}\,,
\end{equation}
    \begin{equation} \label{ShapTyu1b}
l = \pm \frac{x^2 \mp 2 A \sqrt{x} + A^2 }
{\sqrt{x \left( x^2- 3 x \pm 2 A \sqrt{x}\, \right) }}\,,
\end{equation}
    where the upper sign corresponds to the direct orbits (i.e., the orbital
angular momentum of a particle is parallel to the angular momentum of
the black hole), the lower sign corresponds to retrograde orbits,
    \begin{equation} \label{ShapTyu2}
x= \frac{r}{M}\,, \ \ \ \ A= \frac{a}{M}\,, \ \ \ \  l= \frac{L}{M}\,.
\end{equation}
    The circular orbits exist from $r=\infty$ up to
minimal value corresponding to the photon circular orbit~$ r_{\rm ph} $
defined by the roots of the denominator~(\ref{ShapTyu1a}), (\ref{ShapTyu1b})
equal to~\cite{BardeenPressTeukolsky72}
    \begin{equation} \label{ShapTyuPhot}
r_{\rm ph}^\pm = 2 M \left[ 1+ \cos \left( \frac{2}{3} \arccos (\mp A) \right)
\right].
\end{equation}
    The minimal radius of the stable circular orbit is
equal to~\cite{BardeenPressTeukolsky72}
    \begin{equation} \label{ShapTyu4}
x_{\rm ms}^\pm = 3+ Z_2 \mp \sqrt{ (3-Z_1) ( 3+ Z_1 + 2 Z_2) }\,,
\end{equation}
    where
    \begin{equation} \label{ShapTyu5}
Z_1 = 1 + \left( 1\!-\!A^2 \right)^{1/3} \left[ (1\!+\! A)^{1/3} +
(1\!-\!A)^{1/3} \right]\!, \ \ \ \
Z_2 = \sqrt{ 3 A^2 + Z_1^2}\,.
\end{equation}
    The specific energy of the particle on such a limiting stable orbit is\\
$\varepsilon = \sqrt{1- ( 2 / 3 x_{\rm ms})}$.

    The minimal radius of the bounded orbit
(i.e., orbit with $\varepsilon < 1$)
is obtained for~$\varepsilon =1$ (the particle is nonrelativistic at infinity)
and is equal to~\cite{BardeenPressTeukolsky72}:
    \begin{equation} \label{ShapTyu3}
x_{\rm mb}^\pm = 2\left(1+ \sqrt{ 1 \mp A} \, \right) \mp A \,.
\end{equation}
    In this case, we have
    \begin{equation} \label{geodKerr5}
l_{\rm mb}^\pm = \pm 2 \left( 1 + \sqrt{1 \mp A}\, \right).
\end{equation}
    This orbit is nonstable.

    From~(\ref{geodKerr1}), (\ref{ShapTyu1a}), (\ref{ShapTyu1b})
one obtains
    \begin{equation} \label{Zamedl1}
\frac{d t}{d \tau} = \frac{x^{3/2} \pm A }{ \sqrt{ x^{3/2} (x^{3/2} -
3 \sqrt{x} \pm 2 A) }}\,.
\end{equation}
    Due to the fact that $t$ is the time of the observer resting at infinity
from the black hole, $\tau$ is the proper time of the observer moving along
the geodesics, one can consider this value to be ``time dilatation'' for the
corresponding circular orbit.

    For the Schwarzschild black hole $(A=0)$, the time dilatation on the
limiting bounded circular orbit $(x_{\rm mb} =4)$ is equal to 2
(the limiting horizontal dashed-line on Fig.~\ref{DDttL} corresponds
to this value).
    On the minimal stable circular orbit $(x_{\rm ms} =6)$,
the time dilatation is only $\sqrt{2}$.
    However, for circular orbits close to photon orbit $r_{\rm ph}=3M$,
from~(\ref{ShapTyu1a}), (\ref{Zamedl1}) for $A=0$,
using the series expansion in $\varepsilon^{-1}$ one gets
    \begin{equation} \label{Zamedl2}
\frac{d t}{d \tau} = 3 \varepsilon + \frac{1}{6 \varepsilon} +
O\left( \frac{1}{\varepsilon^3} \right), \ \ \ \varepsilon \to \infty,
\end{equation}
    i.e., the time dilatation can be as large as possible.

    Note that in Minkowski space-time one gets from the special
relativity $dt / d \tau = \varepsilon $.
    So in the case of movement around the Schwarzschild black hole, the time
dilatation for relativistic energies of the objects can be enlarged
at most by a factor of three.

    For circular orbits with $ \varepsilon \to \infty $ around the rotating
black holes formulas~(\ref{ShapTyu1a}), (\ref{ShapTyuPhot}), (\ref{Zamedl1})
give
    \begin{equation} \label{Zamedl3}
\frac{d t}{d \tau} \sim \left[ 3 \pm \frac{2 A}{ 2 \cos \! \left(
\frac{\arccos (\mp A)}{3} \right) \mp A} \right] \varepsilon .
\end{equation}
    For direct orbits close to the rapidly rotating black holes with
$ A \to 1 $, $r_{\rm ph}^+ \to M$, it holds
    \begin{equation} \label{Zamedl4}
\frac{d t}{d \tau} \sim \left( \sqrt{ \frac{6}{1-A} } + \frac{1}{3} +
O( \sqrt{1-A} ) \right) \varepsilon ,
\end{equation}
    for retrograde orbits $r_{\rm ph}^- \to 4 M$,
    \begin{equation} \label{Zamedl4o}
\frac{d t}{d \tau} \sim \left( \frac{7}{3} + O(1-A) \right) \varepsilon .
\end{equation}

    The time dilatation on the minimal stable circular orbit of the rapidly
rotating black hole $(A \to 1)$ can be obtained
from~(\ref{ShapTyu1a}), (\ref{ShapTyu4})--(\ref{Zamedl1})
and is equal, for direct orbit $ x_{\rm ms}^+ \to 1$,
    \begin{equation} \label{dtdtlim}
\frac{d t}{d \tau} (x_{\rm ms}^+) = \frac{2^{\,4/3}}{\sqrt{3} \,
(1-A)^{1/3} } \left[ 1 + O ( \sqrt[3]{1-A} ) \right],
\end{equation}
    for the retrograde orbit $x_{\rm ms}^- \approx 9$,
$ d t/ d \tau \approx 13/(6 \sqrt{3})$.
    For minimally bound direct circular orbit, $ x_{\rm mb}^+ \to 1$,
we have
    \begin{equation} \label{dtdtlimlr}
\frac{d t}{d \tau} (x_{\rm mb}^+) = \frac{2}{\sqrt{1-A} }
\left[ 1 + O ( \sqrt{1-A} ) \right].
\end{equation}
    For the retrograde minimally bound orbit:
$x_{\rm mb}^- \approx 3+ 2 \sqrt{2} $, $ d t/ d \tau \approx 3 - \sqrt{2}$.

    Let us give the values of time dilatation on circular orbits for
the black hole with the Thorne's limit for astrophysical black holes
$A=0.998$ (see \cite{Thorne74}):
    \begin{equation} \label{dtdtots}
\frac{d t}{d \tau} (x_{\rm mb}^+) = 43.8 , \ \
\frac{d t}{d \tau} (x_{\rm ms}^+) = 10.8 .
\end{equation}
    For particles with large specific energy rotating in the direction of
rotation of the black hole for orbits close to circular photon orbit one has
    \begin{equation} \label{dtdtThorPh}
A=0.998 \ \ \Rightarrow \\
\frac{d t}{d \tau} \sim 55.12 \, \varepsilon, \ \ \ \varepsilon \to \infty.
\end{equation}
    So the relativistic effect of the time dilatation close to a rotating
black hole may be 55 times larger!

\begin{acknowledgements}
    The author is indebted to Prof. A.A. Grib for the interest in the paper
and useful discussions.
    The research is done in collaboration with Copernicus Center for
Interdisciplinary Studies, Krak\'{o}w, Poland and supported
by the grant from The John Templeton Foundation.
\end{acknowledgements}



\begin{thebibliography}{99}

\bibitem{LL_II}
Landau L.D. and Lifshitz E.M.: {The Classical Theory of Fields}.
Pergamon Press, Oxford (1975)

\bibitem{MTW}
Misner C.W., Thorne K.S. and Wheeler J.A.: {Gravitation.}
Freeman, San Francisco (1973)

\bibitem{Redze}
Regge T.: {Cronache Dell'Universo}.
Boringhieri, Torino (1981)
[Translated into Russian. Mir, Moscow, 1985]

\bibitem{Cherepashchuk}
Cherepashchuk A.M.: Black Holes in the Universe
(in Russian). Vek 2, Fryazino (2005)

\bibitem{GribPavlov2008UFN}
Grib A.A. and Pavlov Yu.V.:
Is it possible to see the infinite future of the Universe when
falling into a black hole?
\href{http://dx.doi.org/10.3367/UFNr.0179.200903d.0279}
{{Uspekhi Fiz. Nauk} {\bf 179}, 279 (2009)}
[English transl.:
\href{http://dx.doi.org/10.3367/UFNe.0179.200903d.0279}
{{Physics -- Uspekhi} {\bf 52}, 257 (2009)}]

\bibitem{Krasnikov08}
Krasnikov S.:
Falling into the Schwarzschild black hole. Important details.
\href{http://dx.doi.org/10.1134/S0202289308040129}
{{Grav. Cosmol.} {\bf 14}, 362 (2008)}

\bibitem{Kruskal60}
Kruskal M.D.:
Maximal extension of Schwarzschild metric.
\href{http://dx.doi.org/10.1103/PhysRev.119.1743}
{Phys. Rev. {\bf 119}, 1743 (1960)}

\bibitem{Szekeres60}
Szekeres G.:
On the singularities of a Riemannian manifold.
{Publ. Mat. Debrecen} {\bf 7}, 285 (1960)

\bibitem{Rees}
Rees M.: {Our Cosmic Habitat.}
Princeton University Press, Princeton (2001)

\bibitem{Schwarzschild16}
Schwarzschild K.:
\"{U}ber das gravitationsfeld eines massenpunktes nach der
Einsteinschen theorie.
{Sitz. Preuss. Akad. Wiss. Berlin}, 189 (1916)

\bibitem{Reissner16}
Reissner H.:
\"{U}ber die eigengravitation des elektrischen feldes nach
der Einsteinschen theorie.
\href{http://dx.doi.org/10.1002/andp.19163550905}
{Ann. Physik {\bf 355}, 106 (1916)}

\bibitem{Nordstrom18}
Nordstr\"{o}m G.:
On the energy of the gravitation field in Einstein's theory.
\href{http://adsabs.harvard.edu/cgi-bin/bib_query?1918KNAB...20.1238N}
{Kon. Nederland. Akad. Wet. Proc. {\bf 20}, 1238 (1918)}

\bibitem{Kottler18}
Kottler F.:
\"{U}ber die physikalischen grundlagen der Einsteinschen gravitationstheorie.
\href{http://dx.doi.org/10.1002/andp.19183611402}
{Ann. Phys. {\bf 361}, 401 (1918)}

\bibitem{Tangherlini63}
Tangherlini F.R.:
Schwarzschild field in $n$ dimensions and the dimensionality of space problem.
\href{http://dx.doi.org/10.1007/BF02784569}
{Nuovo Cimento {\bf 27}, 636 (1963)}

\bibitem{Kerr63}
Kerr R.P.:
Gravitational field of a spinning mass as an example of algebraically
special metrics.
\href{http://dx.doi.org/10.1103/PhysRevLett.11.237}
{Phys. Rev. Lett. {\bf 11}, 237 (1963)}

\bibitem{BoyerLindquist67}
Boyer R.H. and Lindquist R.W.:
Maximal analytic extension of the Kerr metric.
\href{http://dx.doi.org/10.1063/1.1705193}
{J.~Math. Phys. {\bf 8}, 265 (1967)}

\bibitem{Chandrasekhar}
Chandrasekhar S.: {The Mathematical Theory of Black Holes.}
Oxford University Press, Oxford (1983)

\bibitem{BardeenPressTeukolsky72}
Bardeen J.M., Press W.H. and Teukolsky S.A.:
Rotating black holes: locally nonrotating frames, energy extraction,
and scalar synchrotron radiation.
\href{http://adsabs.harvard.edu/full/1972ApJ...178..347B}
{Astrophys.~J. {\bf 178}, 347 (1972)}

\bibitem{Thorne74}
Thorne K.S.:
Disk-accretion onto a black hole. II. Evolution of the hole.
\href{http://dx.doi.org/10.1086/152991}
{Astrophys. J. {\bf 191}, 507 (1974)}

\end{thebibliography}
\end{document}